\documentclass[11pt,twoside]{article}

\usepackage{asp2004}
\usepackage{epsf}
\usepackage{psfig}
\usepackage{lscape}

\begin{document}
\title{The Spitzer Spirals, Bridges, and Tails \\ Interacting Galaxy Survey}   
\author{Beverly J. Smith$^1$, C. Struck$^2$, M. Hancock$^1$, P. Appleton$^3$, W. Reach$^3$, \& 
V. Charmandaris$^4$}   
\affil{$^1$Department of Physics, Astronomy, and Geology, East Tennessee State University, Box 70652, Johnson City TN  37614}
\affil{$^2$Department of Physics and Astronomy, Iowa State University, Ames IA  50011}
\affil{$^3$Spitzer Science Center, California Institute of Technology, Pasadena CA  91125}
\affil{$^4$Department of Physics, University of Crete, 71003 Heraklion, Greece}    

\begin{abstract} 
We present Spitzer mid-infrared images from a survey of three dozen
pre-merger strongly interacting galaxy pairs selected from the Arp Atlas.
The global mid-infrared colors of these galaxies and their tidal tails
and bridges are similar to those of normal spiral galaxies, 
thus this optically selected sample of interacting galaxies does not 
have strongly 
enhanced normalized star formation rates in their disks 
or tidal features.  Despite distortion and disturbance these systems
continue to form stars at a normal rate on average.
The morphology of these galaxies is generally smoother in the shorter
wavelength IRAC bands than at 8 $\mu$m, where dozens of clumps of star
formation are detected.  
\end{abstract}


\section{Introduction}   
Since the issue was raised by \citet{larson78} and \citet{struckmarcell78},
there has been a great deal of interest in how star
formation in galaxies is affected by collisions with other 
galaxies.
Infrared
Astronomical Satellite (IRAS)
observations led to the discovery of galaxies with very high far-infrared
luminosities
\citep{soifer87, smith87}
that are the result of mergers
between equal-mass gas-rich progenitors
\citep{sanders88}.
Later studies showed that spectacular 
enhancement of star formation is the cause
of much of the infrared
emission in major mergers (see review by
\citealp{struck99}).

The question of how star formation is affected in pre-merger 
interacting galaxies
is more difficult to answer.
To date, most studies comparing interacting galaxies to normal
galaxies
have been based on optical data, which suffer extinction,
or IRAS far-infrared data, which provides only a global measurement, and
is complicated by dust heating by older stars.
An alternative tracer of star formation is the mid-infrared
\citep{roussel01, forster04}.
With
the advent of the Spitzer infrared telescope
\citep{werner04}, sensitive higher angular
resolution mid-infrared imaging of galaxies is now possible,
making feasible the detailed study of 
star formation complexes in
interacting systems.
                                                                                
To this end, we have used the Spitzer telescope to observe
a well-defined sample of nearby
interacting
galaxy pairs in the mid-
and far-infrared.
The high sensitivity and good spatial
resolution of Spitzer
make it optimally
suited for studying induced star formation in spiral arms
and
tidal
bridges and tails, and detecting knots of
star formation.

\section{The Samples}

Our interacting galaxy sample was selected from
the Arp Atlas of Peculiar Galaxies
\citep{arp66}, based on the following
criteria:
1) They are relatively isolated binary systems; we
eliminated merger remnants and
close triples and multiple systems in which the galaxies have
similar optical brightnesses (systems with additional smaller angular
size companions were not excluded).
2) They are tidally disturbed.
3) They have
radial velocities less than $<$11,000 km/s (150 Mpc, for H$_{\rm o}$
= 75 km~s$^{-1}$~Mpc$^{-1}$).
4) Their total angular size is $>$ 3$'$, to allow for
good spatial resolution with Spitzer.
5) The angular sizes of the individual
galaxies are $\ge$ 30$''$.
A total of 35 Arp systems fit these criteria.
One of these systems, Arp 297, consists of two pairs at different redshifts,
which are included separately in our sample.
We also include the interacting pair NGC 4567, which fits the above
criteria but is not in the Arp Atlas.  This brings the sample to 37.

Of these 37 systems,
28 were included in our
`Spirals, Bridges, and Tails'
(SB\&T) Guest Observer Cycle 1 Spitzer program.
The remaining 9 galaxies were reserved
as part of various Guaranteed Time or Guest Observer programs.
For completeness,
we also include these additional galaxies.
A few of the SB\&T galaxies were reserved at some wavelengths
and not at others.

As a `control' sample of nearby `normal' galaxies, we started
with
the 75 galaxies in the
Spitzer Legacy `SINGS' project Nearby Galaxies Survey
\citep{kennicutt03, dale05}.
The SINGS sample was selected
to cover a wide range in
parameter space, with a range in Hubble type and luminosity.
Most have angular sizes between 5$'$ and 15$'$.
We excluded from the `normal' sample
SINGS galaxies that are interacting, in close pairs, or in 
compact groups.

\section{Observations}

The galaxies in our SB\&T sample were observed
between November 2004 and November 2005 in
the 3.6, 4.5, 5.8, and 8.0 $\mu$m
broadband
filters of the
Spitzer
Infrared Array Camera (IRAC;
\citealp{fazio04}) and the 24 $\mu$m band
of the Spitzer Multiband Imaging Photometer (MIPS;
\citealp{rieke04}).
A total of 4 $-$ 23 dithered exposures of
12 seconds each were made
per IRAC filter per galaxy, depending upon
the field of view of the system.
For the MIPS observations, we used fixed single observations
with two cycles of 10 sec integration per frame.

\begin{figure}[!ht]
\caption{The 3.6 $\mu$m (left) and 8.0 $\mu$m (right) images of
selected Arp galaxies in our sample.  North is up and east to the left.
The scale bar is 60$''$.
}
\end{figure}

\section{Results and Conclusions}

In general, the galaxies appear smoother in the shorter wavelength
IRAC bands, which are dominated by the older stellar population,
than at 8 and 24 $\mu$m (see Figures 1 $-$ 2).
In the longer wavelength bands, multiple clumps of star formation
are visible, in both the disks and tidal features.
For example, star forming clumps are visible
in our Spitzer images of
the tidal arm/ring of the peculiar galaxy Arp 107 (Figure 1),
and a gradient in the age of the stellar population
is detected along this feature
(see \citealp{smith05}).
In Arp 82, young star forming clumps are also
present in the disk and tidal features (see Hancock et al., this
proceedings).

On average, the Spitzer colors of the disks and tidal features
of the Arp galaxies are similar to those of normal spirals.
Therefore this optically-selected sample of pre-merger interacting
galaxies does not have strongly enhanced star formation compared
to non-interacting galaxies.


\acknowledgements 
We thank the Spitzer and SINGS teams for making this research possible.  
This research was supported by NASA Spitzer grant 1263924, NSF grant
AST-0097616, and NASA LTSA grant NAG5-13079.


\end{document}